\documentclass[onecolumn,secnumarabic,amssymb, nobibnotes, aps, prl,longbibliography,preprint]{revtex4-1}

\usepackage{graphicx,color,amsmath}
\usepackage[normalem]{ulem}

\begin{document}

\title{Adiabatic invariants drive rhythmic human motion in variable gravity}%

\author{N. Boulanger$^1$}
\email[e-mail:]{nicolas.boulanger@umons.ac.be}
\author{F. Buisseret$^{2,3}$}
\email[e-mail:]{buisseretf@helha.be}
\author{V. Dehouck$^1$}
\email[e-mail:]{Victor.DEHOUCK@student.umons.ac.be}
\author{F. Dierick$^{2,4}$}
\email[e-mail:]{dierickf@helha.be}
\author{O. White$^5$}
\email[e-mail:]{olivier.white@u-bourgogne.fr}

\affiliation{$^1$ Service de Physique de l'Univers, Champs et Gravitation, Universit\'{e} de Mons, UMONS  Research
Institute for Complex Systems, Place du Parc 20, 7000 Mons, Belgium }
\affiliation{$^2$ Forme $\&$ Fonctionnement Humain Lab, CeREF, Haute Ecole Louvain en Hainaut, 136 rue Trieu Kaisin, 6061 Montignies sur Sambre, Belgium}
\affiliation{$^3$ Service de Physique Nucl\'{e}aire et Subnucl\'{e}aire, Universit\'{e} de Mons, UMONS Research Institute for Complex Systems, 20 Place du Parc, 7000 Mons, Belgium}
\affiliation{$^4$ Facult\'{e} des Sciences de la Motricit\'e, Universit\'e catholique de Louvain, 1 Place Pierre de Coubertin, 1348 Louvain-la-Neuve, Belgium}
\affiliation{$^5$ Universit\'{e} de Bourgogne INSERM-U1093 Cognition, Action, and Sensorimotor Plasticity, Campus Universitaire, BP 27877, 21078 Dijon, France}
\date{\today}%

\begin{abstract}
Natural human movements are stereotyped. 
They minimise  cost functions that include energy, a natural candidate from mechanical and 
physiological points of view. 
In time-changing environments, however, motor strategies are modified since energy is no longer conserved. 
Adiabatic invariants are relevant observables in such cases, although they have not been 
investigated in human motor control so far. 
We fill this gap and show that the theory of adiabatic invariants 
explains how humans move when gravity varies. 
\end{abstract}

\maketitle

All living organisms experience a constant terrestrial gravitational acceleration, denoted as 1$g$ 
(9.81 m$/s^2$). Gravity, ``the first thing which you don’t think" (A. Einstein), is the most persistent sensory signal in the brain. However, the sensory experiences it generates lack the clear phenomenology of an identifiable 
stimulus event that characterises sound, sight and even taste. Critically, gravity influences human 
behaviour more pervasively than any other sensory signal. 
Exposure to Earth-discrepant gravity -- as during spaceflight -- leads to dramatic structural 
and functional changes in the human physiology, including alterations in the
cardiovascular \cite{Aubert16}, neural \cite{White16} and musculoskeletal systems \cite{Lang17}. Nowadays the cerebellum appears to be a major structure in gravity perception \cite{macneilage18}, but we still have no complete understanding of how the brain processes gravity to plan and control actions.

Recent neurocomputational approaches explain behaviour by a mixture of feedback and 
feedforward mechanisms, conceptualised by internal models \cite{Kawato99}: 
the brain plans an action using available sensory information and makes predictions about 
the consequences of that action in the environment. Any mismatch between this prediction 
and the information conveyed by feedback will yield a prediction error used to improve other actions. 
This mechanism drives motor adaptation. 
On Earth, gravity is immutable and plays a primary role in minimising prediction errors by providing a strong prior reference. 

What is the best way to fundamentally address the role of gravity in motor control? 
One radical approach consists in challenging the brain by changing a feature of the environment 
that is never supposed to change: gravity itself. 
Our original approach is to assess the impact of time-changing gravity on rhythmic biological 
motion from a purely mechanical vantage point, thereby providing further insights into the fundamental 
representation of gravity that shapes motor actions. 
Living organisms are extraordinarily more complex than a simple point-particle body. 
It is not at all obvious that the actions of a minded human being can be reduced to a standard, 
simple Lagrangian. Lifting a glass of water off a table requires estimating its weight 
to adjust the grasping force accordingly. Drinking half of its content with a straw while 
the glass rests on the table does not, however, allow the brain to program a smaller grasping force, 
more adapted to the lighter glass \cite{Nowak}. 
Explicit knowledge of the simplest change in object dynamics is not sufficient to update internal models. Therefore, our working hypothesis is that human actions comply 
with the behaviour of a mechanical system, even if subject to a slowly changing environment, 
like a slowly varying gravitational field.

In Mechanics, the most robust way to track the adaptation of a dynamical system to a slow change in the 
external conditions is through the study of adiabatic invariants and their related action-angle variables describing the system \cite{L&L}. 
An adiabatic invariant determines a property of a system that stays approximately constant when external changes occur slowly. Despite their power in revealing constraints 
on complex dynamical systems, adiabatic invariants have been poorly investigated in biomechanics. 
For instance, in arm rhythmic motion, the changes in frequency ($df$) occurring during a 
one-dimensional periodic motion are correlated with changes in energy ($dE$) \cite{turvey:1996} 
such that the action variable 
\begin{equation}\label{Idef}
I=\frac{1}{2\pi}\frac{dE}{df}
\end{equation}
is constant. Action-angle coordinates are usually adopted when the Hamiltonian 
does not depend explicitly on time. 
The present work goes beyond previous approaches by immersing participants in a time-dependent 
gravitational environment where the action variables are not necessarily constant unless 
the changes in time are adiabatic.

The action-angle variables appeared in the context of classical mechanics in order to study the 
integrability of dynamical systems with finitely many degrees of freedom. 
Such systems are said to be \textit{integrable} if the Hamilton-Jacobi equation describing them 
is completely separable. In the early sixties, the famous Kolmogorov-Arnold-Moser theorem 
--- see \cite{Dumas} for a very interesting book telling the history behind this theorem --- 
brought back the action-angle variables on the scene of classical Mechanics in order to 
characterise chaotic Hamiltonian systems. Since then and with the seminal works of Nekhoroshev 
\cite{nekhoroshev1971behavior,nekhoroshev1977exponential} their importance has never faded out. 
When a Hamiltonian $H(P_\alpha,Q^\alpha)\,$, $\alpha=1,\ldots, n\,$, is integrable and leads 
to bounded trajectories in phase space, action variables may be defined as follows, 
in terms of a set of phase-space coordinates that separates the Hamiltonian:
\begin{equation}\label{Id0}
    I_\alpha = \frac{1}{2\pi}\oint_{\Gamma_\alpha} 
    P_\alpha\, dQ^\alpha\ ,
\end{equation}
where $\Gamma_\alpha$ is the projection of the bounded trajectory in the plane $(P_\alpha,Q^\alpha)$ 
for fixed $\alpha\,$. Once the Hamilton-Jacobi equation is separated in the variables $(Q^\alpha,P_\alpha)\,$, 
on the solution of Hamilton’s canonical equations each momentum variable $P_\alpha$  will depend only on 
its canonically conjugate variable $Q^\alpha$ and on the initial conditions. 
The action variables give all the conserved quantities of the dynamical system under study, 
as certified by the Bour-Liouville theorem.

If the Hamiltonian is time-dependent and slowly varying in comparison with 
the typical period of a cycle, then the action variables are slowly varying too. 
They are called adiabatic invariants \cite{L&L,henrard,jose}  and may be used in a 
wide range of applications such as in electromagnetism \cite{tennyson86}, 
plasma physics \cite{notte93} and cosmology \cite{cotsakis98}. 
Previous works in biomechanics showed the invariance of the action variable 
when experimental conditions are time-independent \cite{kugler:1990,turvey:1996,kadar:1993}. 
To the best of our knowledge, this concept has never been applied to human motion 
in time-varying environments. Our approach can reveal the important and otherwise hidden quantities on which the brain relies to plan actions. Advances in this field can potentially not be reached with other, more classical, methods that rest on energy conservation \cite{Alexander}. We therefore designed an experimental set up in which external factors are time-dependent. It is described in the next paragraph.

\begin{figure}
\includegraphics[width=\columnwidth]{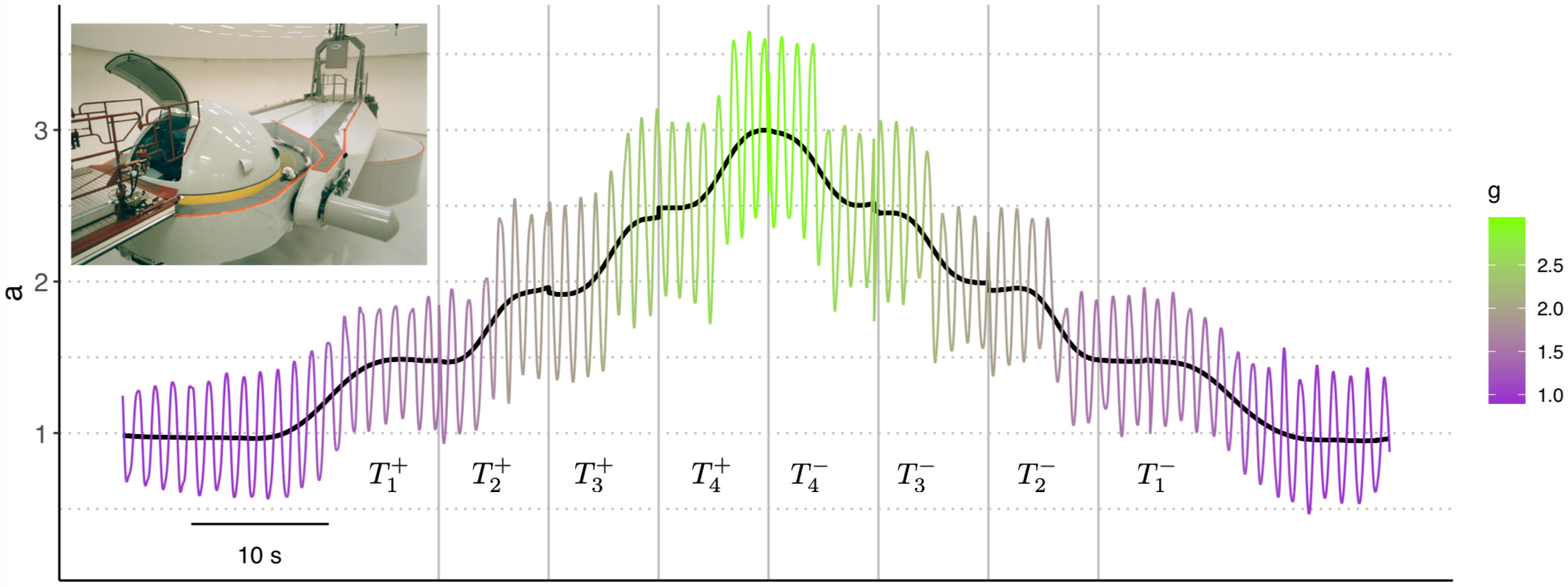} 
\caption{Typical plot of raw data recorded by the accelerometer (coloured line) during a single session of centrifugation (inset). The wireless test object (mass of 0.13 kg) embedded an accelerometer that measured combined gravitational and kinematic accelerations along the object’s long axis (AIS326DQ, range 30m$/$s$^2$, accuracy $\pm 0.2$m$/$s$^2$). The acceleration signal was sampled at a frequency of 120Hz. The black line depicts local gravity. All accelerations are expressed in units of $g=9.81$ m$/$s$^2$. The plateau phases are shown for the first and last transitions. For the other transitions, plateau phases and rest periods are not displayed for the sake of clarity but are replaced by vertical lines.
}\label{fig0}
\end{figure}

Six right-handed male participants ($40.1\pm7.2$ years old) took part in two centrifugation 
sessions at QinetiQ’s Flight Physiological Centre in Link\"{o}ping, Sweden. 
The centrifuge was controlled to deliver specific $g(t)$-profiles. 
The real-time control of the orientation of the gondola ensured alignment of local gravity 
with the long body axis (Fig. \ref{fig0} inset). One session of centrifugation consisted in 
a ramp up followed by a ramp down $g(t)$-profile for 180s. 
There were two equivalent sessions separated by a five-minute break bringing the centrifuge back to idle position. The initial 1$g$ phases (idle) lasted for 27.4s. 
Then, the system generated 1.5$g$, 2$g$, 2.5$g$, 3$g$, 2.5$g$, 2$g$, 1.5$g$ and 1$g\,$. 
Each phase lasted 18.4s and transitions lasted 1.6s (average rate of 0.31$g$/s), 
except for the first and last ones. 
We label a given transition by $T^\pm_n$ where it is meant that $g(t)$ goes 
from the value $(n+1) g/2$ to the value $(n+1+\eta) g/2$, with $\eta=\pm 1\,$. The increasing (decreasing) gravitational transitions correspond to $\eta=+1$ ($-1)\,$. 
In both cases, $n\in\{1,2,3,4\}\,$. 
The first decreasing-$g$ series is $T_4^-$  while the last one is $T_1^-\,$ (Fig. \ref{fig0}). 
A medical flight doctor assessed the participant’s health status before the experiment. 
The protocol was reviewed and approved by the Facility Engineer from the Swedish Defence Material 
Administration (FMV) and an independent medical officer. The experiment was overseen 
by a qualified medical officer. The study was conducted in accordance with the Declaration 
of Helsinki (1964). All participants gave informed and written consent prior to the study.
A similar protocol was used in a previous study where the human centrifuge is described 
in detail \cite{white18}.

Participants performed upper arm rhythmic movements about the elbow at a free, 
comfortable pace only during the transitions between gravitational environments, to limit fatigue. 
When prompted by a GO signal, the participant started to perform the movement while holding 
an object embedding an accelerometer. The elbow remained in contact with the support. 
The upper arm produced movements of about 30$^{\rm o}$ with the horizontal. 
When the operator announced the STOP signal, the participant gently let the object touch 
the support again while still securing it with his hand. 
A schematic representation of raw data (acceleration vs time) of one session for one 
subject is displayed in Fig. \ref{fig0}.

\begin{figure}
\includegraphics[width=\columnwidth]{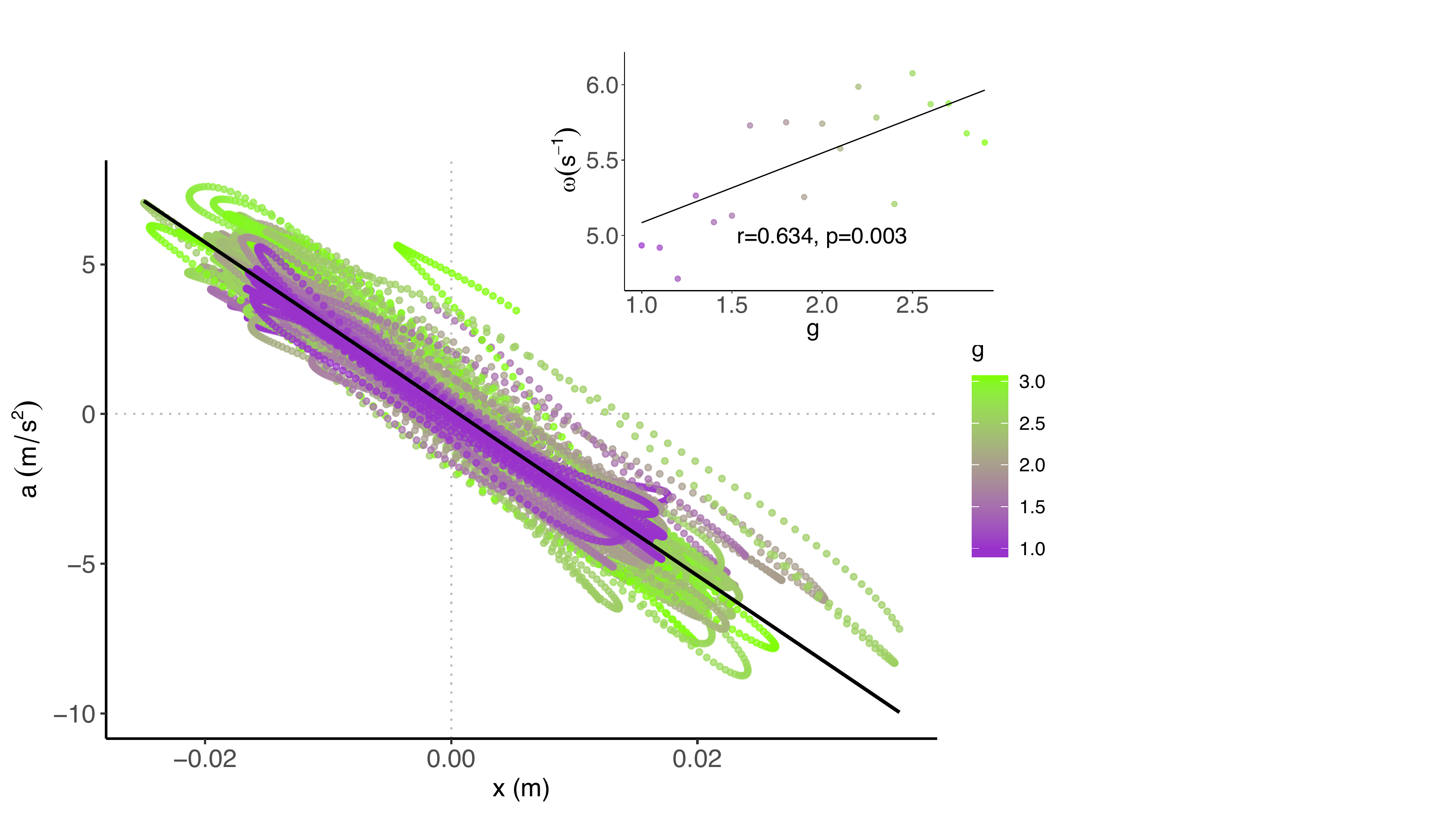} 
\caption{Typical plot of acceleration versus position for the test object during one centrifugation session, same participant as Fig. \ref{fig0} (coloured points). A global linear regression is shown (solid line).  The inset quantifies the significant linear relationship between $\omega$ and $g$
. Dots result from a fit of the form (\ref{eqho}) by bins of 0.1 $g$.
}\label{figHO}
\end{figure}

\begin{figure}
\includegraphics[width=\columnwidth]{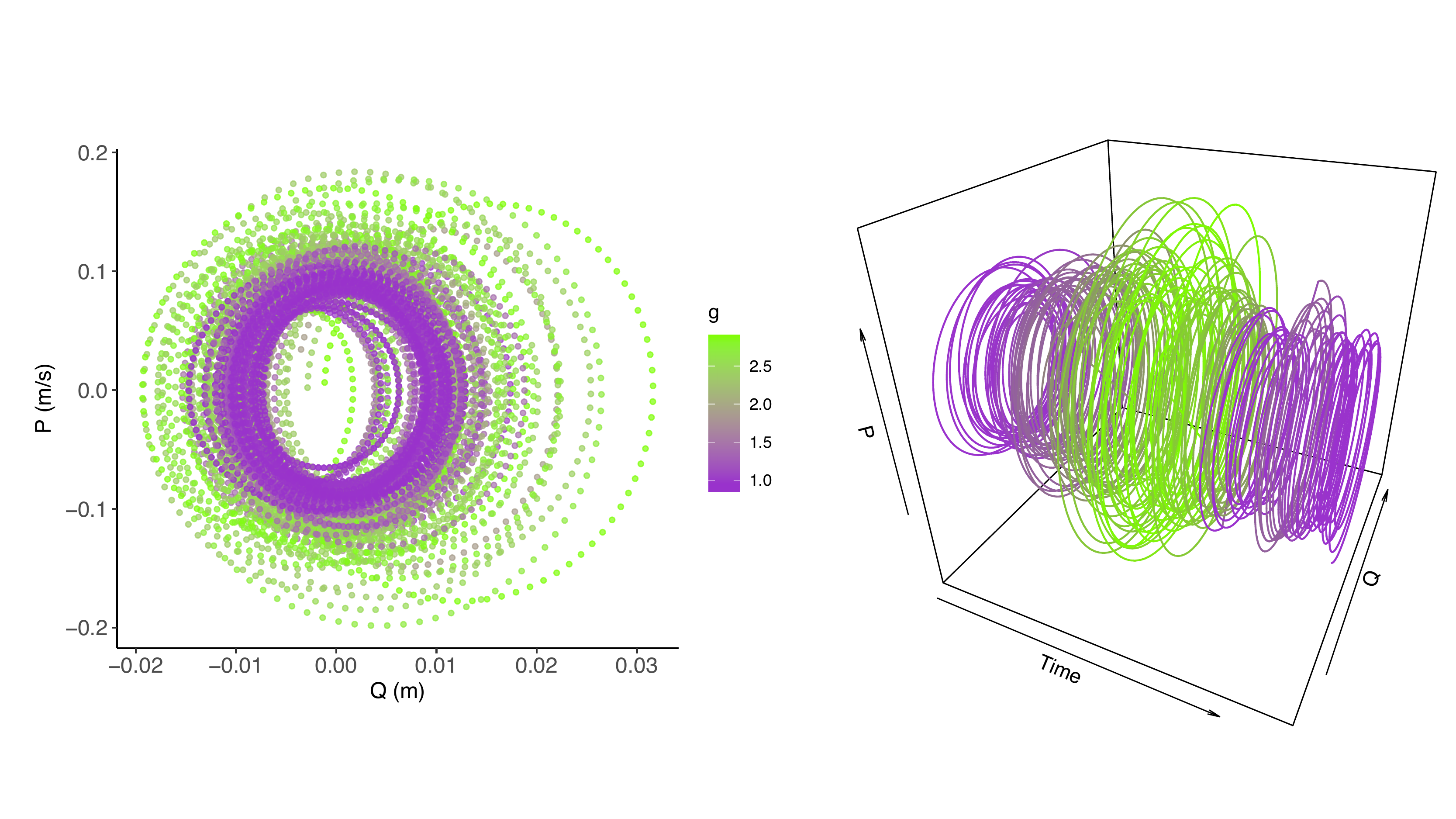} 
\caption{Left panel: Typical phase-space plot of the test object trajectory during one centrifugation session, same participant as Fig. \ref{fig0}. Right panel: Same data but the consecutive cycles are now unfolded along the time dimension. 
}\label{fig1}
\end{figure}

Accelerations $a(t)$ were numerically integrated and linearly detrended after subtraction 
of $g(t)$ to yield the object’s speed and position $x(t)$. The link
\begin{equation}\label{eqho}
a=-\omega^2\ x    
\end{equation}
is observed for all participants within a given transition (96 time series): averaged Pearson's correlation coefficient between $a$ and $x$ is indeed equal to $-0.82\pm0.1$. 
A typical plot is shown in Fig. \ref{figHO}. In average, $\omega=6.3$ Hz leading to a typical 
period T$=0.99$ s. Hence, we are on safe grounds to assume that the dynamics of the test object along the body axis is compatible with that of a harmonic oscillator, 
\textit{i.e.}, with a Hamiltonian of the form
\begin{eqnarray}\label{Hdef}
    H&=&\frac{P^2}{2}+\frac{1}{2}\omega(t)^2 Q^2\;, ~~{\rm with}\quad P =\dot Q\quad {\rm and}\ Q=x\;.
\end{eqnarray}
Figure \ref{fig1} depicts a typical phase-space of a complete centrifugation session. Elliptic cycles are clearly visible and are the consequence of the harmonic-oscillator dynamics. The area of these ellipses is slowly changing with $g$ as expected from adiabatic invariant’s theory. Action-angle coordinates $(I,\phi)$ may be defined through the standard definition \cite{L&L}
\begin{equation}\label{PQIphi}
Q=\sqrt{\frac{2I}{\omega}}\sin\phi\ ,\ P=\sqrt{2I\omega}\cos\phi
\end{equation}
and their equations of motion read 
\begin{equation}\label{eqIphi}
    \dot I=-\frac{\dot \omega}{\omega}I\, \cos 2\phi\ , \ \dot \phi=\omega+\frac{\dot \omega}{2\omega}\sin 2\phi \ .
\end{equation}

The parameter of model (\ref{Hdef}) is the function $\omega(g(t))\,$. 
Careful inspection of experimental data let us conclude that $\omega(g(t))$ is compatible with a weakly 
increasing linear shape, see Fig. \ref{figHO} inset. 
Hence we assume
\begin{equation}\label{omegadef}
    \omega(t)=\varpi \left(1 +\frac{\epsilon}{g}\, g(t)\right)\ 
\end{equation}
and we will perform computations up to first order in $\epsilon$ through the rest of the paper. Equation (\ref{omegadef}) is justified physiologically: muscle stiffness increases with gravitational acceleration to account for the larger motor commands required to perform the same movement. This leads to a modified frequency and $\epsilon>0\,$. Similarly, muscle stiffness should decrease from normal to microgravity.

Let us now focus on a given transition $T^{\pm}_n$. Equation (\ref{omegadef}) can be adapted to the peculiar shape of $g(t)$ imposed during the centrifugation:
\begin{eqnarray}\label{model_centrif}
    \omega_n(t)&=&\varpi_n \left(1 +\epsilon\, s(t)\right)\ ,
    \quad 
\varpi_n = \omega_0\Big(1+\frac{\epsilon}{2}\, 
(n-\tfrac{1}{2})\,\Big)\ ,\nonumber \\ 
s(t) &=& \frac{\eta}{4}\, \sin(\Omega t)\ , ~\eta =\pm1 \ , \quad 
{\rm with} \ t\in\left[-\frac{\pi}{2\Omega} ,\frac{\pi}{2\Omega} \right]\ .
\end{eqnarray}
We have shown in \cite{Boulanger:2018tue} that $I(t)$ and $\phi(t)$ can be analytically computed at order $\epsilon$ from 
Eq. (\ref{eqIphi}) when $g(t)$ is of trigonometric form. 
This gives
\begin{eqnarray}\label{eqIphi2}
I(t)&=& \bar I \left[1-\epsilon\, \eta\, \frac{\Omega}{16}\,\left( \frac{1}{\omega^+}\sin[2(\omega^+t +\alpha)]+(+\leftrightarrow -)\right)\right]\ , \nonumber \\
\phi(t)&=&\alpha+\varpi_n t-\epsilon\,\eta\, \frac{\omega_0}{4\Omega}\,\cos(\Omega t) \nonumber \\
&&-\epsilon\, \eta\, \frac{\Omega}{32}\left( \frac{1}{\omega^+}\cos[ 2(\omega^+ t+\alpha)]+(+\leftrightarrow -)]\right)\ , 
\end{eqnarray}
%
with $\omega^\pm=\omega_0\pm\frac{\Omega}{2}$ and $\omega_0 > \Omega\,$.

The action variable takes a simpler form when $P=0$, \textit{i.e.} for $t_k$ such that
\begin{eqnarray}
    \phi(t_k) &=&(2k+1)\pi/2 =:\phi_k\ , \qquad k\in\mathbb{Z}\ ,
\end{eqnarray}
see Eq. (\ref{PQIphi}). The analytical shape of the times $t_k$ such that $\phi(t_k)=\phi_k$
may be complicated but since our goal is the computation of $I(t_k)$, it is sufficient to work with the lowest order solution $t_k=\frac{\phi_k-\alpha}{\omega_0}$, 
leading to 
\begin{equation}
    I(t_k)=\bar I \left(1-\epsilon \;\frac{\Omega^2}{4\omega_0^2-\Omega^2}\;s(t_k)\right).
\end{equation}

For a given transition $T^\pm_n\,$, $g(t)/g=\tfrac{n+6}{2} + s(t)\,$. 
Hence, $I(t_k)=A_{n,\eta}+B\, g(t_k)$, where $A_{n,\eta}$ and $B$ are real constants, and where $B=dI/dg$ does not depend on $n$ and $\eta\,$. It allows us to append the transitions and get an affine relation between $I(t_k)$ and $g(t_k)$ during the whole centrifugation session:
\begin{eqnarray}\label{model}
    I(t_k)&=:& I_0+I_1\, g(t_k)\ , 
\end{eqnarray}
with $I_0\in\mathbb{R}^+$ and $I_1\in\mathbb{R}\,$. The shift in $I(t)$ predicted by Eqs. (\ref{eqIphi2}) and (\ref{model}) extend previous results 
obtained in Ref. \cite{Kulsrud:1957zz} where an analytical shape is obtained for $I(t)$ with arbitrary 
$\omega(t)$ provided that the latter is not $C^\infty\,$.

\begin{figure}
\includegraphics[width=\columnwidth]{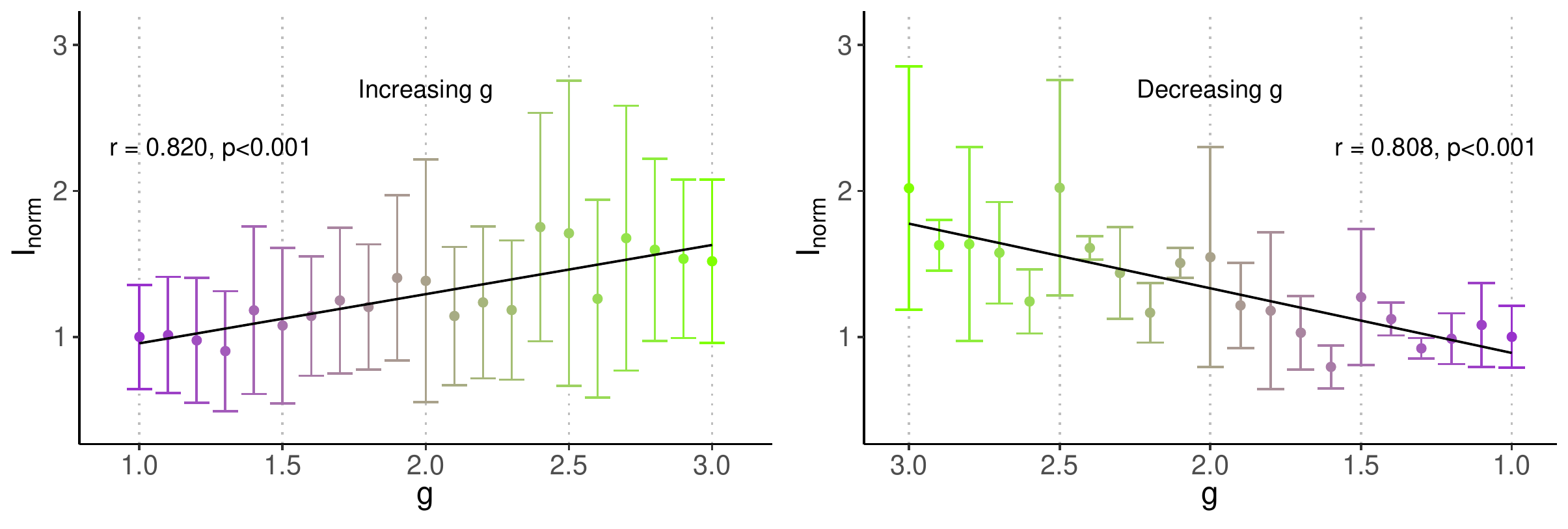} 
\caption{Mean values (and 1 SD error bars) of the adiabatic invariant $I_{\rm norm}$ per bin, normalised 
to the 1$g$ value, versus $g(t)$. Significant linear regressions of the experimental data are depicted as 
a solid black line together with their Pearson’s correlation coefficients and $p$-values. 
The left panel presents data in the ascending $g(t)$ phase and the right panel presents data in the 
descending $g(t)$ phase. Note that in the descending phase, the horizontal axis is decreasing in 
order to provide a continuous and chronological reading of the evolution of $I_{\rm norm}$. 
}\label{fig2}
\end{figure}

We have computed phase-space trajectories of all participants in both centrifugation sessions. 
It is therefore possible to compute the action variable as a function of time. 
Indeed, Eq. (\ref{Id0}) can be rewritten as $ I(t)=\int^{t^*}_{t} \dot Q^2\, dt\,$, with $t^*$ 
the end of the phase-space cycle starting at $t$. The instant $t^*>t$ is such that the distance between 
the points $(Q(t),P(t))$ and $(Q(t^*),P(t^*))$ in phase space is minimal and the difference $t^*- t$ is as 
close as possible to T. Once the action variables $I(t)$ are known, the times $t_k$ such that $P(t_k)=0$ 
are computed as well as the action variables $I(t_k)$. Continuous values $I(t_k)$ of all participants and 
all trials are finally discretised into 0.1 $g$-bins ranging from 1 to 3 $g$. Each bin contains between 14 
and 23 data points. Average values and standard deviations (SD) of $I$ normalised to the 1$g$ value are 
finally displayed in Fig. \ref{fig2}.

The adiabatic invariant exhibits a strong and significant positive ($I_1>0$) linear relationship with gravity 
both in the increasing and decreasing phases (Fig. \ref{fig2}). According to Eq. (\ref{Idef}), it shows a 
higher energetic cost in high gravity for a given change in frequency, which is expected since raising the 
test object by a height $\Delta h$ has a potential energetic cost of order $m g \Delta h$.

Despite this overall coherent dependence of $I$ over $g\,$, we observed asymmetries in the slopes $I_1$ 
(Eq. \ref{model}) between ascending and descending phases. To quantify this effect, we ran a 
2-way repeated measures ANOVA with factors session (1 or 2) and phase (increasing or decreasing). 
This analysis shows that the slope $I_1$ is significantly larger in the increasing phase than in the 
decreasing phase ($I_1=0.296\pm0.306 > 0.523\pm0.219$, $p=0.037$). This asymmetry was not influenced 
by session ($p=0.130$). The fact that the adiabatic invariant exhibits a stronger dependence on $g$ 
in the ascending phases is an interesting observation. 
It indicates that the adiabatic invariant may be modulated more by vestibular and/or proprioceptive gains 
and re-adjustments of central pattern generators (CPGs) and/or cerebellum activities during that phase. 
At a spinal cord level indeed, rhythmic movements in mammals are organised by network of interneurons 
and motor neurons called CPGs \cite{cpg}.
Here, movements of the forearm were generated by CPGs located at cervico-thoracic level \cite{zehr}. The observation of rapid adaptation of rhythmic forearm movements suggests that vestibular 
and proprioceptive feedback are the major source of information used by CPGs to ensure adjustments 
to altered gravity, especially when it increases and becomes more demanding for the control of the task. 
Interestingly, CPGs receive both inputs from proprioceptive afferents and vestibular pathways. At a supraspinal level, the cerebellum could be the structure integrating the variations of gravity \cite{macneilage18}, eventually leading to behaviours compatible with adiabatic invariants. 

The variability of $I_{\rm norm}$ at a given $g$ is globally lower 
in the decreasing than in the increasing-$g$ phase as can  be seen from the error bars.  It suggests habituation takes place because the decreasing-$g$ phase always follows the increasing-$g$ one. The higher variability during the increasing phase is consistent with the realisation of a movement in a new situation. 
During the decreasing phase, motor learning achieved in the previous phase made 
it possible to induce a gradual reduction of variability in order to optimise the movement 
patterns that are compatible with a simple harmonic oscillator.

In summary, participants show a spontaneous adaptation of their motion that 
is compatible with the expectation of a simple harmonic oscillator with weakly gravity-dependent frequency. 
Their adaptation is assessed by the computation of adiabatic invariants, whose experimental behaviour 
versus $g$ comply with our model’s prediction. We hypothesise that the main biological 
receptors of time-changing gravity are proprioceptors, such as muscle spindles and Golgi tendon 
organs that are known to give constant feedback to the CPGs. Adiabatic invariants may thus put realistic constraints on the choices made by spinal 
and supraspinal nervous structures among an infinite number of possible solutions to a given problem, 
\textit{i.e.}, the motion of our test object in the present case. 
Such ``hidden'' constraints in voluntary motion may be of interest in domains such as 
rehabilitation and robotics.
 
Future works might go beyond the harmonic oscillator description of the effective dynamics but still in a 
phase-space based formalism. 
As shown in \cite{Boulanger:2018tue}, adiabatic invariants can be computed in the case of 
higher-derivative Hamiltonians of Pais-Uhlenbeck type. Such Hamiltonians could describe rhythmic 
motions with several frequencies and discrete movements through, \textit{e.g.}, minimal jerk 
models \cite{Hogan}. We are currently investigating how our model can be generalised by analysing 3D trajectories performed during parabolic flight, 
therefore also including the very particular case of an absence of gravity \cite{white08,tocome}.

\textit{Acknowledgements}
This research was supported by the European Space Agency (ESA) in the framework of the 
Delta-G Topical Team (4000106291/12/NL/VJ), the ``Institut National de la Sant\' e et de la Recherche M\' edicale'' (INSERM) and the ``Conseil G\' en\' eral de Bourgogne'' (France) and by the “Centre National d'Etudes Spatiales” grant 4800000665 (CNES). We thank E. Ferr\`e for inspiring parts of this manuscript

\bibliographystyle{hunsrt}
\bibliography{biblio_adiabatic}

\begin{thebibliography}{10}

\bibitem{Aubert16}
A.E. Aubert et~al.
\newblock Towards human exploration of space: the theseus review series on
  cardiovascular, respiratory, and renal research priorities.
\newblock {\em npj Microgravity}, 2:16031, 2016.

\bibitem{White16}
O.~White et~al.
\newblock Towards human exploration of space: the theseus review series on
  neurophysiology research priorities.
\newblock {\em npj Microgravity}, 2:16023, 2016.

\bibitem{Lang17}
T.~Lang, J.J.W.A. Van~Loon, S.~Bloomfield, L.~Vico, A.~Chopard, J.~Rittweger,
  A.~Kyparos, D.~Blottner, I.~Vuori, R.~Gerzer, and P.R. Cavanagh.
\newblock Towards human exploration of space: the theseus review series on
  muscle and bone research priorities.
\newblock {\em npj Microgravity}, 3:8, 2017.

\bibitem{macneilage18}
P.R. MacNeilage and S.~Glasauer.
\newblock Gravity perception: The role of the cerebellum.
\newblock {\em Current Biology}, 28, 2018.

\bibitem{Kawato99}
M.~Kawato.
\newblock Internal models for motor control and trajectory planning.
\newblock {\em Curr. Opin. Neurobiol.}, 9:718--727, 1999.

\bibitem{Nowak}
D.A. Nowak and J.~Hermsdörfer.
\newblock Sensorimotor memory and grip force control: does grip force
  anticipate a self-produced weight change when drinking with a straw from a
  cup?
\newblock {\em Eur. J. Neurosci}, 18, 2003.

\bibitem{L&L}
L.~Landau and E.~Lifchitz.
\newblock {\em {Physique th\'{e}orique Tome 1 : M\'{e}canique}}.
\newblock E. MIR, Moscow, 1988.

\bibitem{turvey:1996}
M.T. Turvey, K.G. Holt, J.~Obusek, et~al.
\newblock {Adiabatic transformability hypothesis of human locomotion}.
\newblock {\em Biol. Cybern.}, 74(107):107--115, 1996.

\bibitem{Dumas}
H.S. Dumas.
\newblock {\em {The KAM story: a friendly introduction to the content, history,
  and significance of classical Kolmogorov-Arnold-Moser theory}}.
\newblock World Scientific, Hackensack, NJ, Apr 2014.

\bibitem{nekhoroshev1971behavior}
N.N. Nekhoroshev.
\newblock Behavior of hamiltonian systems close to integrable.
\newblock {\em Functional Analysis and Its Applications}, 5(4):338--339, 1971.

\bibitem{nekhoroshev1977exponential}
N.N. Nekhoroshev.
\newblock An exponential estimate of the time of stability of nearly-integrable
  hamiltonian systems.
\newblock {\em Uspekhi Matematicheskikh Nauk}, 32(6):5--66, 1977.

\bibitem{henrard}
J.~Henrard.
\newblock {\em The Adiabatic Invariant in Classical Mechanics}, pages 60--73.
\newblock Dessy, 1998.

\bibitem{jose}
J.V. Jose and E.J Saletan.
\newblock {\em {Classical dynamics: a contemporary approach}}.
\newblock Cambridge Univ. Press, Cambridge, 1998.

\bibitem{tennyson86}
J.~L. Tennyson, John~R. Cary, and D.~F. Escande.
\newblock Change of the adiabatic invariant due to separatrix crossing.
\newblock {\em Phys. Rev. Lett.}, 56:2117--2120, May 1986.

\bibitem{notte93}
J.~Notte, J.~Fajans, R.~Chu, and J.~S. Wurtele.
\newblock Experimental breaking of an adiabatic invariant.
\newblock {\em Phys. Rev. Lett.}, 70:3900--3903, Jun 1993.

\bibitem{cotsakis98}
S.~Cotsakis, R.~L. Lemmer, and P.~G.~L. Leach.
\newblock Adiabatic invariants and mixmaster catastrophes.
\newblock {\em Phys. Rev. D}, 57:4691--4698, Apr 1998.

\bibitem{kugler:1990}
P.N. Kugler, M.T. Turvey, R.C. Schmidt, and L.D. Rosenblum.
\newblock {Investigating a Nonconservative Invariant of Motion in Coordinated
  Rhythmic Movements}.
\newblock {\em Ecological Psychology}, 2(2):151--189, 1990.

\bibitem{kadar:1993}
E.E. Kadar, R.C. Schmidt, and M.T. Turvey.
\newblock {Constants underlying frequency changes in biological rhythmic
  movements}.
\newblock {\em Biol. Cybern.}, 68:421--430, 1993.

\bibitem{Alexander}
R.~McN. Alexander.
\newblock A minimum energy cost hypothesis for human arm trajectories.
\newblock {\em Biol. Cybern.}, 76, 1997.

\bibitem{white18}
O.~White, J.-L. Thonnard, Ph. Lef\`evre, and J.~Hermsd\"orfer.
\newblock Grip force adjustments reflect prediction of dynamic consequences in
  varying gravitoinertial fields.
\newblock {\em Frontiers in Physiology}, 9:131, 2018.

\bibitem{Boulanger:2018tue}
N.~Boulanger, F.~Buisseret, F.~Dierick, and O.~White.
\newblock {Higher-derivative harmonic oscillators: stability of classical
  dynamics and adiabatic invariants}.
\newblock {\em Eur. Phys. J.}, C79(1):60, 2019, 1811.07733.

\bibitem{Kulsrud:1957zz}
R.M. Kulsrud.
\newblock {Adiabatic Invariant of the Harmonic Oscillator}.
\newblock {\em Phys. Rev.}, 106:205--207, 1957.

\bibitem{cpg}
E.~Marder and D.~Bucher.
\newblock Central pattern generators and the control of rhythmic movements.
\newblock {\em Current Biology}, 11, 2012.

\bibitem{zehr}
E.P. Zehr et~al.
\newblock Possible contributions of cpg activity to the control of rhythmic
  human arm movement.
\newblock {\em Can. J. Physiol. Pharmacol.}, 82, 2004.

\bibitem{Hogan}
T.~Flash and N.~Hogan.
\newblock The coordination of arm movements: an experimentally confirmed
  mathematical model.
\newblock {\em J. Neurosci}, 5, 1985.

\bibitem{white08}
O.~White et~al.
\newblock Altered gravity highlights central pattern generator mechanisms.
\newblock {\em J Neurophysiol}, 100, 2008.

\bibitem{tocome}
N.~Boulanger, F.~Buisseret, V.~Dehouck, F.~Dierick, and O.~White.
\newblock Rhythmic motion in hyper- and micro-gravity: The role of adiabatic
  invariants in motor strategy.
\newblock {\em in preparation}, 2019.

\end{thebibliography}

\end{document}